\documentclass[11pt]{article}
\pdfoutput=1 
\usepackage{graphicx}
\usepackage{setspace}
\begin{document}
\noindent
{\Large Blazar spectral variability as explained by a twisting inhomogeneous jet}

\bigskip\noindent
{C.~M.~Raiteri$^{ 1}$,
M.~Villata$^{1}$,
J.~A.~Acosta-Pulido$^{2,3}$,
I.~Agudo$^{4}$,
A.~A.~Arkharov$^{5}$,
R.~Bachev$^{6}$,
G.~V.~Baida$^{7}$,
E.~Ben\'itez$^{8}$,
G.~A.~Borman$^{7}$,
W.~Boschin$^{9,2,3}$,
V.~Bozhilov$^{10}$,
M.~S.~Butuzova$^{ 7}$,
P.~Calcidese$^{11}$,
M.~I.~Carnerero$^{1}$,
D.~Carosati$^{12,9}$,
C.~Casadio$^{13,4}$,
N.~Castro-Segura$^{3,14}$,
W.-P.~Chen$^{15}$,
G.~Damljanovic$^{16}$, \\
F.~D'Ammando$^{17,18}$,
A.~Di~Paola$^{19}$,
J.~Echevarr\'ia$^{8}$,
N.~V.~Efimova$^{5}$, \\
Sh.~A.~Ehgamberdiev$^{20}$,
C.~Espinosa$^{8}$,
A.~Fuentes$^{4}$,
A.~Giunta$^{19}$,
J.~L.~G\'omez$^{4}$,
T.~S.~Grishina$^{21}$,
M.~A.~Gurwell$^{22}$,
D.~Hiriart$^{8}$,
H.~Jermak$^{23}$,
B.~Jordan$^{24}$,
S.~G.~Jorstad$^{25,21}$,
M.~Joshi$^{25}$,
E.~N.~Kopatskaya$^{21}$,
K.~Kuratov$^{26,27}$, \\
O.~M.~Kurtanidze$^{28,29,30,31}$,
S.~O.~Kurtanidze$^{28}$,
A.~L\"ahteenm\"aki$^{32,33,34}$, \\
V.~M.~Larionov$^{21,5}$,
E.~G.~Larionova$^{21}$,
L.~V.~Larionova$^{21}$,
C.~L\'azaro$^{2,3}$,
C.~S.~Lin$^{15}$,
M.~P.~Malmrose$^{25}$,
A.~P.~Marscher$^{25}$,
K.~Matsumoto$^{35}$,
B.~McBreen$^{36}$,
R.~Michel$^{8}$,
B.~Mihov$^{6}$,
M.~Minev$^{10}$,
D.~O.~Mirzaqulov$^{20}$,
A.~A.~Mokrushina$^{21,5}$,
S.~N.~Molina$^{4}$,
J.~W.~Moody$^{37}$,
D.~A.~Morozova$^{21}$,
S.~V.~Nazarov$^{7}$,
M.~G.~Nikolashvili$^{28}$,
J.~M.~Ohlert$^{38,39}$,
D.~N.~Okhmat$^{7}$,
E.~Ovcharov$^{10}$,
F.~Pinna$^{2,3}$,
T.~A.~Polakis$^{40}$,
C.~Protasio$^{2,3}$,
T.~Pursimo$^{41}$,
F.~J.~Redondo-Lorenzo$^{2,3}$,
N.~Rizzi$^{42}$,
G.~Rodriguez-Coira$^{2,3}$,
K.~Sadakane$^{35}$,
A.~C.~Sadun$^{43}$,
M.~R.~Samal$^{15}$,
S.~S.~Savchenko$^{21}$,
E.~Semkov$^{6}$,
B.~A.~Skiff$^{44}$,
L.~Slavcheva-Mihova$^{6}$,
P.~S.~Smith$^{45}$,
I.~A.~Steele$^{23}$,
A.~Strigachev$^{6}$,
J.~Tammi$^{32}$,
C.~Thum$^{46}$,
M.~Tornikoski$^{32}$,
Yu.~V.~Troitskaya$^{21}$,
I.~S.~Troitsky$^{21}$,
A.~A.~Vasilyev$^{21}$,
and O.~Vince$^{16}$
}

\bigskip\noindent
{\footnotesize
$^{ 1}$INAF, Osservatorio Astrofisico di Torino, I-10025 Pino Torinese, Italy                                                                                \\
$^{ 2}$Instituto de Astrofisica de Canarias (IAC), La Laguna, E-38200 Tenerife, Spain                                                                        \\
$^{ 3}$Departamento de Astrofisica, Universidad de La Laguna, La Laguna, E-38205 Tenerife, Spain                                                             \\
$^{ 4}$Instituto de Astrof\'{\i}sica de Andaluc\'{\i}a (CSIC), E-18080 Granada, Spain                                                                        \\
$^{ 5}$Pulkovo Observatory, 196140 St.\ Petersburg, Russia                                                                                                   \\
$^{ 6}$Institute of Astronomy and NAO, Bulgarian Academy of Sciences, 1784 Sofia, Bulgaria                                                                   \\
$^{ 7}$Crimean Astrophysical Observatory RAS, P/O Nauchny, 298409, Russia                                                                                    \\
$^{ 8}$Instituto de Astronom\'ia, Universidad Nacional Aut\'onoma de M\'exico, M\'exico                                                                      \\
$^{ 9}$INAF, TNG Fundaci\'on Galileo Galilei, E-38712 La Palma, Spain                                                                                        \\
$^{10}$Department of Astronomy, Faculty of Physics, University of Sofia, BG-1164 Sofia, Bulgaria                                                             \\
$^{11}$Osservatorio Astronomico della Regione Autonoma Valle d'Aosta, I-11020 Nus, Italy                                                                     \\
$^{12}$EPT Observatories, Tijarafe, E-38780 La Palma, Spain                                                                                                  \\
$^{13}$Max-Planck-Institut f\"ur Radioastronomie, D--53121, Bonn, Germany                                                                                    \\
$^{14}$School of Physics \& Astronomy, University of Southampton, Southampton, SO17 1BJ, UK                                                                  \\
$^{15}$Graduate Institute of Astronomy, National Central University, Jhongli City, Taoyuan County 32001, Taiwan                                              \\
$^{16}$Astronomical Observatory, 11060 Belgrade, Serbia                                                                                                      \\
$^{17}$Dip.\ di Fisica e Astronomia, Universit\`a di Bologna, I-40129 Bologna, Italy                                                                         \\
$^{18}$INAF, Istituto di Radioastronomia, I-40129 Bologna, Italy                                                                                             \\
$^{19}$INAF, Osservatorio Astronomico di Roma, I-00040 Monte Porzio Catone, Italy                                                                            \\
$^{20}$Ulugh Beg Astronomical Institute, Maidanak Observatory, Tashkent, 100052, Uzbekistan                                                                  \\
$^{21}$Astronomical Institute, St.\ Petersburg State University, 198504 St.\ Petersburg, Russia                                                              \\
$^{22}$Harvard-Smithsonian Center for Astrophysics, Cambridge MA 02138 USA                                                                                   \\
$^{23}$Astrophysics Research Institute, Liverpool John Moores University, Liverpool L3 5RF, UK                                                               \\
$^{24}$School of Cosmic Physics, Dublin Institute For Advanced Studies, Ireland                                                                              \\
$^{25}$Institute for Astrophysical Research, Boston University, Boston, MA 02215, USA                                                                        \\
$^{26}$NNLOT, Al-Farabi Kazakh National University, Almaty, Kazakhstan                                                                                       \\
$^{27}$Fesenkov Astrophysical Institute, Almaty, Kazakhstan                                                                                                  \\
$^{28}$Abastumani Observatory, Mt. Kanobili, 0301 Abastumani, Georgia                                                                                        \\
$^{29}$Engelhardt Astronomical Observatory, Kazan Federal University, Tatarstan, Russia                                                                      \\
$^{30}$Landessternwarte, Zentrum für Astronomie der Universität Heidelberg, 69117 Heidelberg, Germany                                                      \\
$^{31}$Center for Astrophysics, Guangzhou University, Guangzhou, 510006, China                                                                               \\
$^{32}$Aalto University Mets\"ahovi Radio Observatory, FI-02540 Kylm\"al\"a, Finland                                                                         \\
$^{33}$Aalto University Dept of Electronics and Nanoengineering, FI-00076 Aalto, Finland                                                                     \\
$^{34}$Tartu Observatory, 61602 T\~{o}ravere, Estonia                                                                                                        \\
$^{35}$Astronomical Institute, Osaka Kyoiku University, Osaka, 582-8582, Japan                                                                               \\
$^{36}$UCD School of Physics, University College Dublin, Dublin 4, Ireland                                                                                   \\
$^{37}$Department of Physics and Astronomy, Brigham Young University, Provo, UT 84602, USA                                                                   \\
$^{38}$Michael Adrian Observatorium, Astronomie Stiftung Trebur, 65468 Trebur, Germany                                                                       \\
$^{39}$University of Applied Sciences, Technische Hochschule Mittelhessen, 61169 Friedberg, Germany                                                          \\
$^{40}$Command Module Observatory, Tempe  AZ, USA                                                                                                            \\
$^{41}$Nordic Optical Telescope, E-38700 Santa Cruz de La Palma, Santa Cruz de Tenerife, Spain                                                               \\
$^{42}$Osservatorio Astronomico Sirio, I-70013 Castellana Grotte, Italy                                                                                      \\
$^{43}$Department of Physics, University of Colorado Denver, CO, 80217-3364 USA                                                                              \\
$^{44}$Lowell Observatory, Flagstaff  AZ, USA                                                                                                                \\
$^{45}$Steward Observatory, University of Arizona, Tucson, AZ, USA                                                                                           \\
$^{46}$Instituto de Radio Astronom\'ia Milim\'etrica, E-18012 Granada, Spain                                                                                 \\
 }
 
\bigskip\noindent
{\bf Blazar emission is dominated by non-thermal radiation from a relativistic jet pointing toward us, therefore undergoing Doppler beaming$^1$. 
This is responsible for flux enhancement and contraction of the variability time scales, so that most blazars appear as luminous sources characterized by noticeable and fast flux changes at all frequencies.
The mechanisms producing their unpredictable variability are debated and include injection, acceleration and cooling of particles$^2$, with possible intervention of shock waves$^{3,4}$ or turbulence$^5$. 
Changes in the viewing angle of the emitting knots or jet regions have also been suggested to explain flaring events$^{6,7,8,9,10}$ or specific properties such as intraday variability$^{11}$, quasi-periodicities$^{12,13}$, or the delay of radio flux variations relative to optical changes$^{14}$. However, such a geometric interpretation has not been universally accepted because alternative explanations based on changes of physical conditions can also work in many cases$^{15,16}$. 
Here we report the results of optical-to-radio monitoring of the blazar CTA~102 by the Whole Earth Blazar Telescope Collaboration and show that the observed long-term flux and spectral variability is best explained by an inhomogeneous, curved jet that undergoes orientation changes. 
We propose that magnetohydrodynamic instabilities$^{17}$ or rotation of a twisted jet$^{6}$ cause different jet regions to change their orientation and hence their relative Doppler factors. 
In particular, the recent extreme optical outburst (six magnitudes) occurred when the corresponding jet emitting region acquired a minimum viewing angle.} 

\bigskip\noindent
CTA~102 belongs to the flat-spectrum radio quasar (FSRQ) subclass of blazars. Its redshift $z=1.037$ corresponds to a luminosity distance of about 7000 Mpc (assuming a flat Universe and a Hubble constant $H_0=70 \, \rm km \, s^{-1} \, Mpc^{-1}$).

The Whole Earth Blazar Telescope (WEBT) Collaboration started to monitor the source multiwavelength behaviour in 2008. Data up to 2013 January were included in [\ref{lar16}].
In Methods we give some details on the observations and in Extended data Fig.\ \ref{webt} we show the optical and near-infrared light curves in 2013--2017 built with data from 39 telescopes in 28 observatories.

A period of relatively low activity has recently been interrupted by a sudden rise of the source brightness in late 2016, with a jump of 6--7 magnitudes with respect to the minima in the optical and near-infrared bands.
 
\begin{figure}
\center\includegraphics[width=\linewidth]{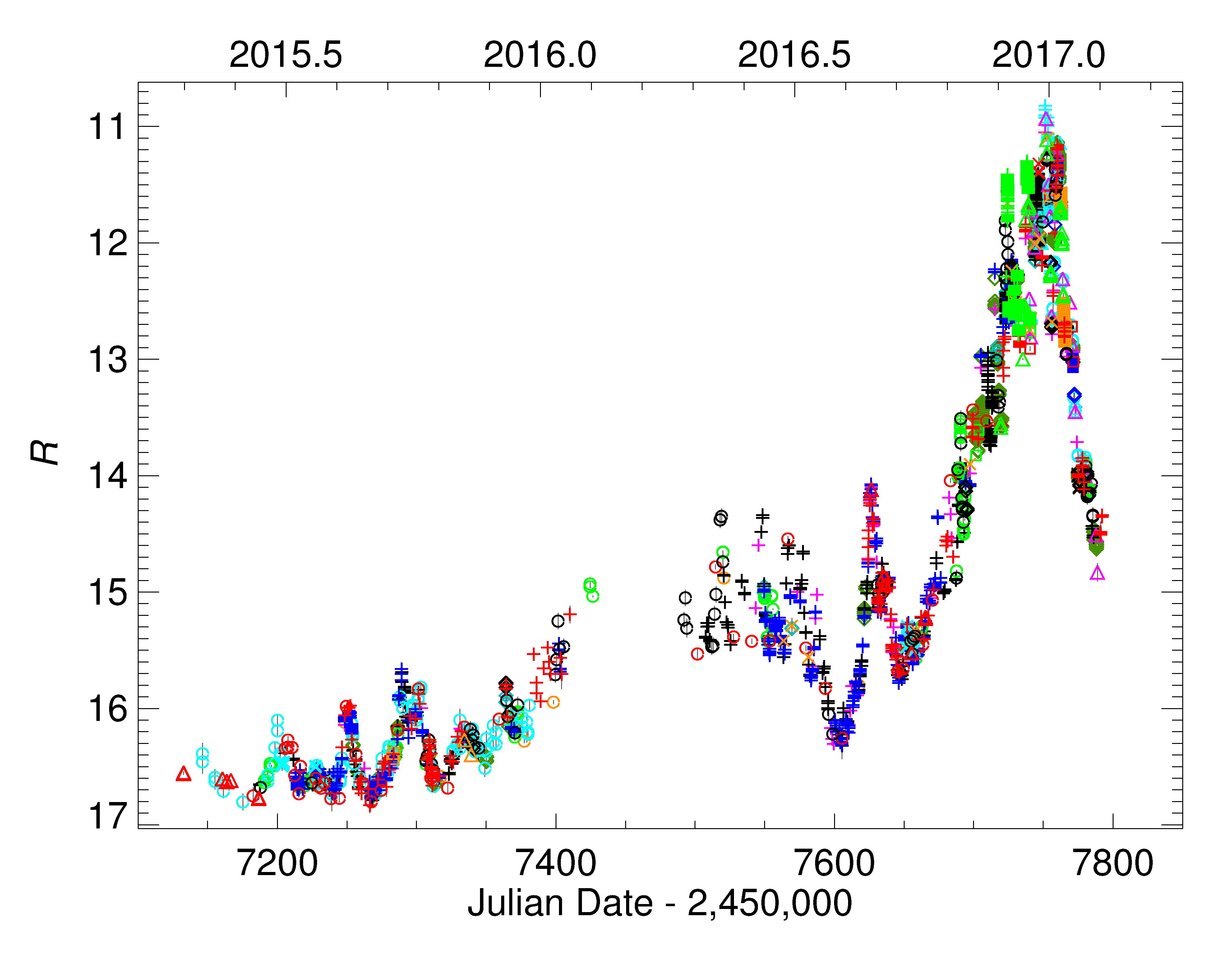}
\caption{{\bf The observed optical light curve of CTA~102 in the last two observing seasons.} $R$-band magnitudes are shown as a function of the Julian Date. Different colours and symbols distinguish the various telescopes contributing to the WEBT campaign. Bars represent 1 s.d. measure errors. The peak of the 2016--2017 outburst was observed on December 28 and implies a $\sim 6$ mag brightness increase with respect to the faintest state.}
\label{outburst}
\end{figure}

The peak of the outburst was observed on December 28, with an $R$-band magnitude of $10.82 \pm 0.04$ (see Fig.\ \ref{outburst}), corresponding to a luminosity ($k$-corrected but without a beaming correction) of $\log (\nu L_\nu) = 48.12$ [$\rm erg \, s^{-1}$]. 
This represents the most luminous optical blazar state ever detected. 
The previous record was held by the FSRQ 3C~454.3, that during the big outburst$^{\ref{vil06}}$ observed in 2005 reached $\log (\nu L_\nu) = 47.54$.

Inspection of the light curves reveals that the variability amplitude is larger in the near-infrared than in the optical band. This is a consequence of the source emitting not only synchrotron radiation from the jet, but also thermal radiation from the
accretion disc$^{\ref{rai14}}$ that feeds the super-massive black hole of the active galactic nucleus (AGN). The more stable light from the disc makes a larger contribution to the overall source flux at optical wavelengths than in the near-IR.
Further evidence of disc thermal radiation comes from the analysis of colour indices and spectroscopic data (see Methods and Extended Data Figs.\ \ref{colori} and \ref{spettri}).

\begin{figure}
\center\includegraphics[width=11cm]{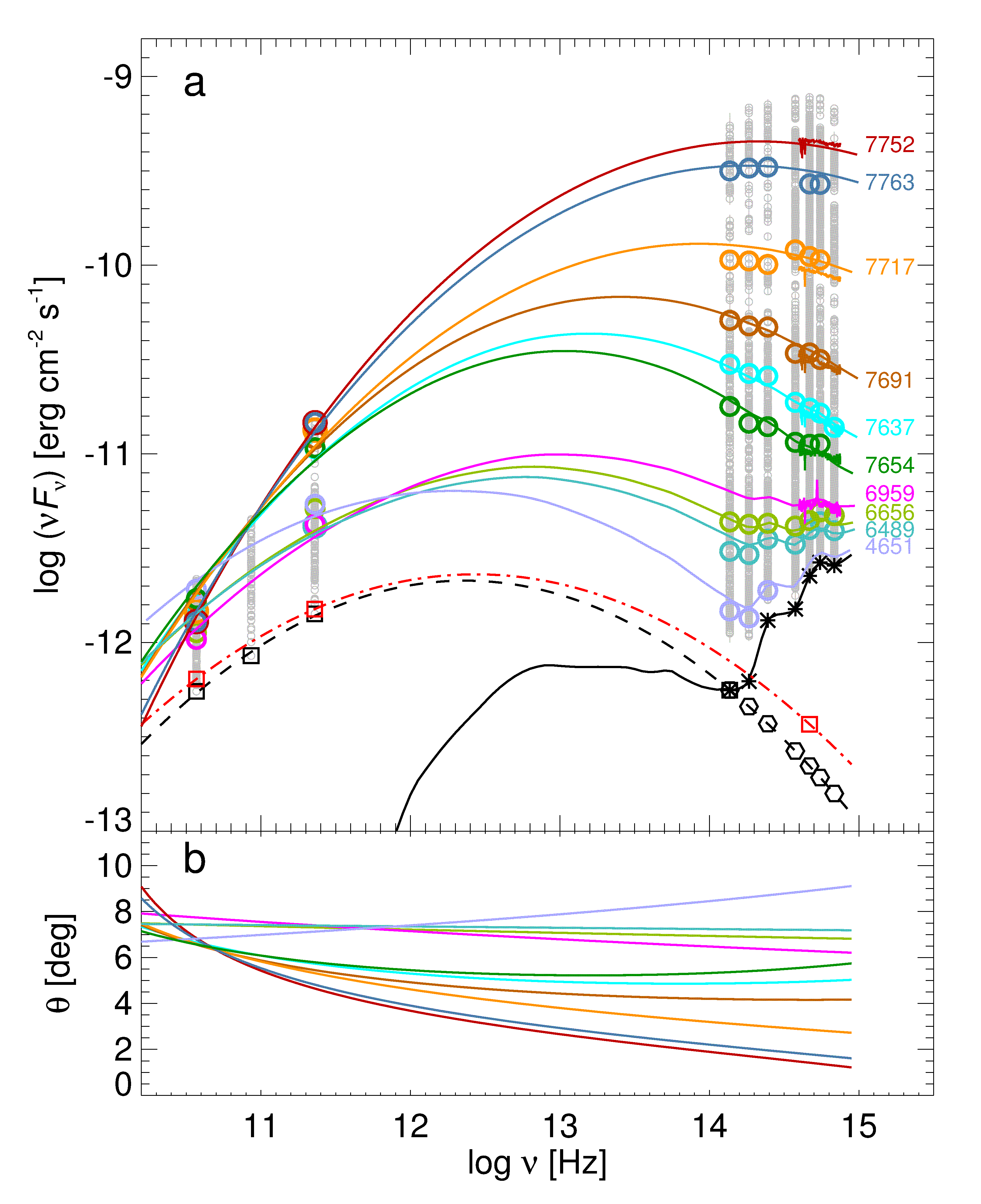}
\caption{{\bf Spectral energy distributions of CTA~102 and orientation of the jet emitting regions}. 
{\bf a}, Small grey circles highlight the observed variability ranges. 
The black dashed line represents the putative minimum synchrotron SED, black squares (hexagons) the minimum synchrotron flux densities fitted (derived). 
Black solid line and asterisks show the thermal emission model and its contributions to the near-infrared and optical bands. The dot-dashed red line represents the base-level synchrotron SED for the geometrical interpretation. Coloured big circles and lines display observed data and spectra and model predictions for selected epochs (labelled with $\rm JD-2450000$). Measure errors (1 s.d.) are smaller than the symbols size. 
{\bf b}, The viewing angles of the jet emitting region producing the (bulk of the) radiation at the frequency $\nu$ at the epochs in the upper panel.}
\label{sed}
\end{figure}

In order to analyse the jet synchrotron emission, we must first model the thermal component, usually referred to as the ``big blue bump" (BBB). Besides the disc radiation, the BBB includes the contribution of emission lines from the broad-line region of the AGN; in particular, Mg~II and $\rm H \alpha$ lines, redshifted to the optical $V$ and near-infrared $J$ bands, respectively. Details on the BBB modelling are given in Methods. In summary, we built the spectral energy distribution (SED) of a putative synchrotron minimum brightness state and then subtracted it from the flux minima in all optical and near-infrared bands to get the BBB contribution. We also added a dust torus emission component in the mid--far infrared, as dust emission in CTA~102 has been detected with the IRAS$^{\ref{imp88}}$ and Spitzer$^{\ref{mal11}}$ satellites. The results are shown in Fig.\ \ref{sed}.

\begin{figure}
\center\includegraphics[width=11cm]{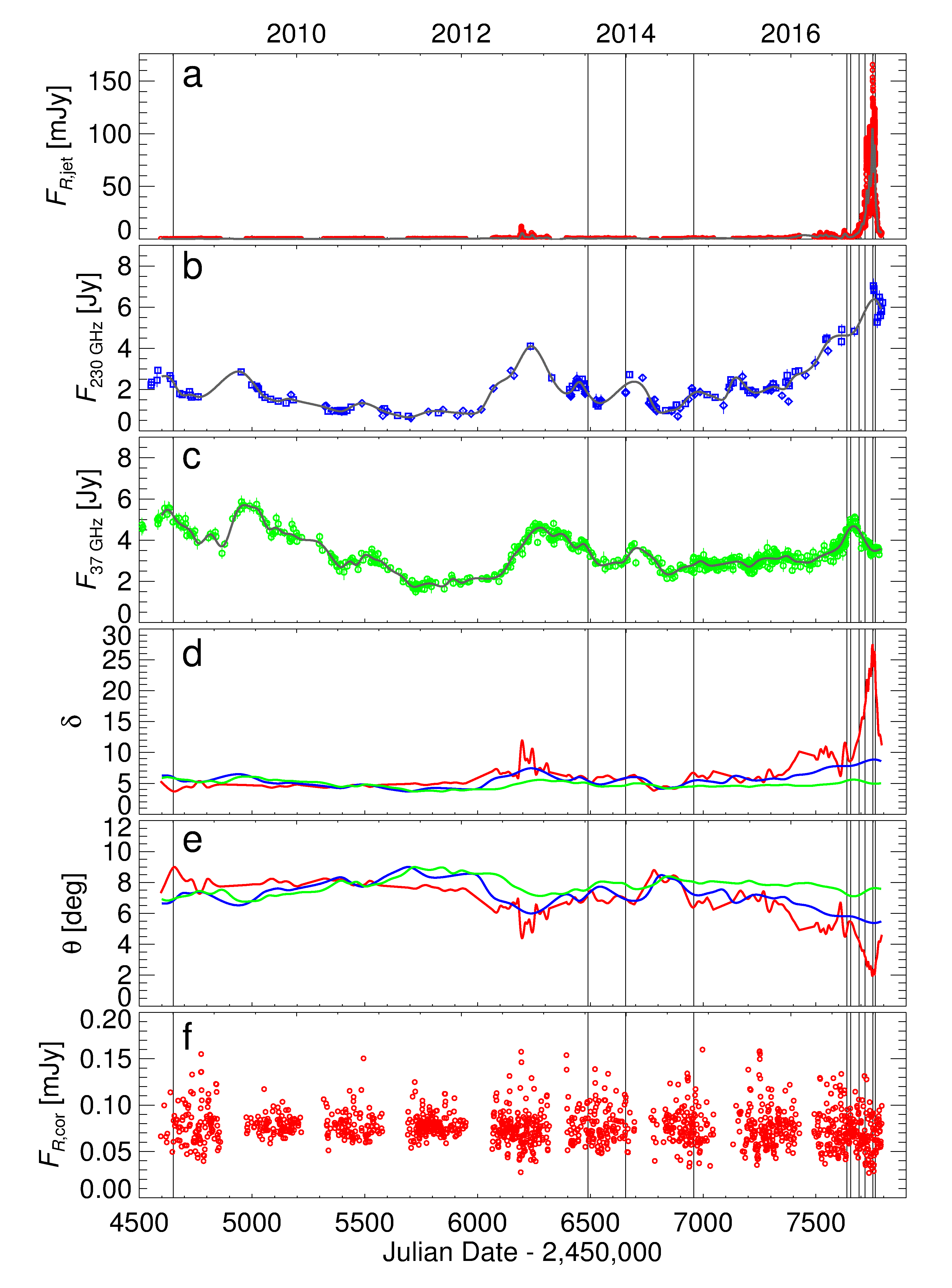}
\caption{{\bf Multifrequency behaviour of the CTA~102 jet emission in 2008--2017.}
{\bf a}-{\bf c}, The $R$-band, 230 GHz and 37 GHz light curves contain 10462, 170 and 576 flux density points, respectively. Bars represent 1 s.d. measure errors. Grey solid lines are cubic spline interpolations through the binned light curves. {\bf d}-{\bf e}, The trend of the Doppler factor $\delta$ and viewing angle $\theta$ of the optical (red), 230 GHz (blue) and 37 GHz (green) jet emitting regions according to the geometrical model. {\bf f}, The $R$-band flux densities corrected for the variable beaming effect; they were obtained for a constant $\delta=\delta_{\rm base}$. The vertical lines indicate the epochs considered in Fig.\ \ref{sed}.}
\label{radop}
\end{figure}

Having a model for the thermal contribution to the source flux, we subtract it from the near-infrared and optical flux densities and get the jet synchrotron flux. 
Fig.\ \ref{radop} shows the optical $R$-band, millimetric (230 GHz) and radio (37 GHz) light curves in the period 2008--2017. The jet optical flux density ranges from 0.047 to 166 mJy, with a maximum flux ratio of more than 3500.
We note that both the 2012 flare and the 2016--2017 outburst were accompanied by radio activity, but the flux ratios at the peaks of the two events are very different in the various bands. 
Moreover, in 2012, peaks at lower frequencies were following those at higher frequencies, as is often observed in blazars$^{\ref{vil09}}$. In contrast, the last optical outburst was preceded by activity at 37 GHz. 
Fig.\ \ref{radop} also shows that the high flux densities registered at 37 GHz in 2008--2009 correspond to a ``quiescent" optical state.
This complex optical-radio correlation suggests that the emission in these two bands is produced in different jet regions. 
Explaining the multiwavelength light curves in terms of intrinsic processes would require very different physical conditions along the jet at various epochs. 
Therefore, we look for an alternative scenario and try to see if the observed source behaviour can rather be ascribed to orientation changes in the jet.

In the light curves, especially in the optical and near-infrared bands, we can distinguish fast flares superimposed on a long-term trend.
We adopt cubic spline interpolations through the binned data to represent the long-term behaviour of the well-sampled light curves in the $R$ band and at 230 and 37 GHz.
For the radio and mm light curves, which are characterised by smooth variations, we use a fixed time bin of 30 days. In the optical, the fast flares are more rapid and pronounced when the source is brighter. This is what is expected if the long-term trend is due to a variable Doppler factor $\delta$, which affects both the flux variation amplitudes and the time scales.
In Methods we summarize the basic concepts of the relativistic beaming theory and verify the contraction of variability time scales during the brightest optical states. Accordingly, we adopted a variable bin size, which goes from an initial value of 24 days in low flux states to 3.4 days in the most dramatic phases of the 2016--2017 outburst.
Nevertheless, the dispersion (root mean square) of the optical flux densities around the spline in the various observing seasons goes from 0.03 mJy, to 1.1 mJy at the time of the 2012 flare, and to 18 mJy during the 2016--2017 outburst. This confirms that fast flares are amplified during high flux states, strongly indicating that the long-term flux changes are likely due to Doppler factor variations. 

If we accept this,
we can trace the behaviour of $\delta$ in time (see Fig.\ \ref{radop}) at the three reference frequencies.
We must take into account that, because of the Doppler beaming, what we observe at a given frequency is emitted by the source at a lower frequency.
To correct for this effect we use the relativistic invariant $F_\nu/\nu^2$ (see e.g.\ [\ref{urr95}]), where $F_\nu$ is the flux density at frequency $\nu$.
We build a base-level synchrotron spectrum for the long-term flux variations by fitting a log-parabolic model to the spline minima at 37 and 230 GHz and in the $R$ band (see Fig.\ \ref{sed}).
This is what we assume to be the source SED subject to the minimum Doppler factor $\delta_{\rm base}$.
Starting from here, for each observed $F_\nu$ we look for the corresponding frequency $\nu_{\rm base}$ in the base-level spectrum so that $F_\nu (\nu)/\nu^2=F_{\nu,{\rm base}} (\nu_{\rm base})/\nu_{\rm base}^2$. Once $\nu_{\rm base}$ is found, we can infer the Doppler factor as $\delta=\delta_{\rm base} \, (\nu/\nu_{\rm base})$.
The trends of $\delta$ shown in Fig.\ \ref{radop} were obtained assuming typical values$^{\ref{sav10}}$ for the bulk Lorentz factor $\Gamma=20$ and for the maximum viewing angle $\theta_{\rm max}=9^\circ$, so that $\delta_{\rm base}=3.7$. Other choices for $\Gamma$ and $\theta_{\rm max}$ do not alter the general results. 
Data constrain the ratio $\delta(t)/\delta_{\rm base}$, while the choice of $\theta_{\rm max}$ constrains $\Gamma$ to yield a reasonable $\theta_{\rm min}$. In the light of what is known for blazars, values of $\theta_{\rm max}$ between about 5$^\circ$ and 15$^\circ$, corresponding to values of $\Gamma$ between 35 and 10, are possible.

The Doppler factor depends on the bulk Lorentz factor and on the viewing angle. While changes of $\Gamma$ both along the jet (see Methods) and in time are in principle possible, they would require large differential accelerations/decelerations of the bulk flow in the various jet regions to explain the extreme flux changes in CTA~102. Instead, we favour Doppler factor variations caused by orientation changes, which is also supported by the development of non-axisymmetric instabilities in magnetohydrodynamic (MHD) jet simulations$^{\ref{mig10}}$ and by the VLBI observations (though on much larger, parsec scales) of swirling jets$^{\ref{bri17}}$ or helical jet structures$^{\ref{per12}}$, also in the case of CTA 102$^{\ref{fro13}}$. 

Having $\delta (t)$ and a guess for $\Gamma$, we can then derive the viewing angle as a function of time.
This is shown in Fig.\ \ref{radop} for the three reference bands.
Flux enhancements are seen at a given frequency when the corresponding jet emitting region becomes better aligned with the line of sight. This happens in particular in the optical at the time of the 2016--2017 outburst.

If we now correct the flux densities for the variable $\delta$ effect (see Methods), 
we obtain what we would observe in the case that the jet had a constant orientation in time, and the same for all its emitting regions, i.e.\ a constant $\delta$.
The residual variability corresponds to the fast flares that are possibly caused by intrinsic, energetic processes. The fast flares show similar amplitudes over the whole 2008--2017 period. The dispersion in the various observing seasons is reduced to a factor 2 (0.009--0.022 mJy), compared to a factor 600 if energetic processes within the jet exclusively explain the observed variability.

\begin{figure}
\center\includegraphics[width=9cm,angle=-90]{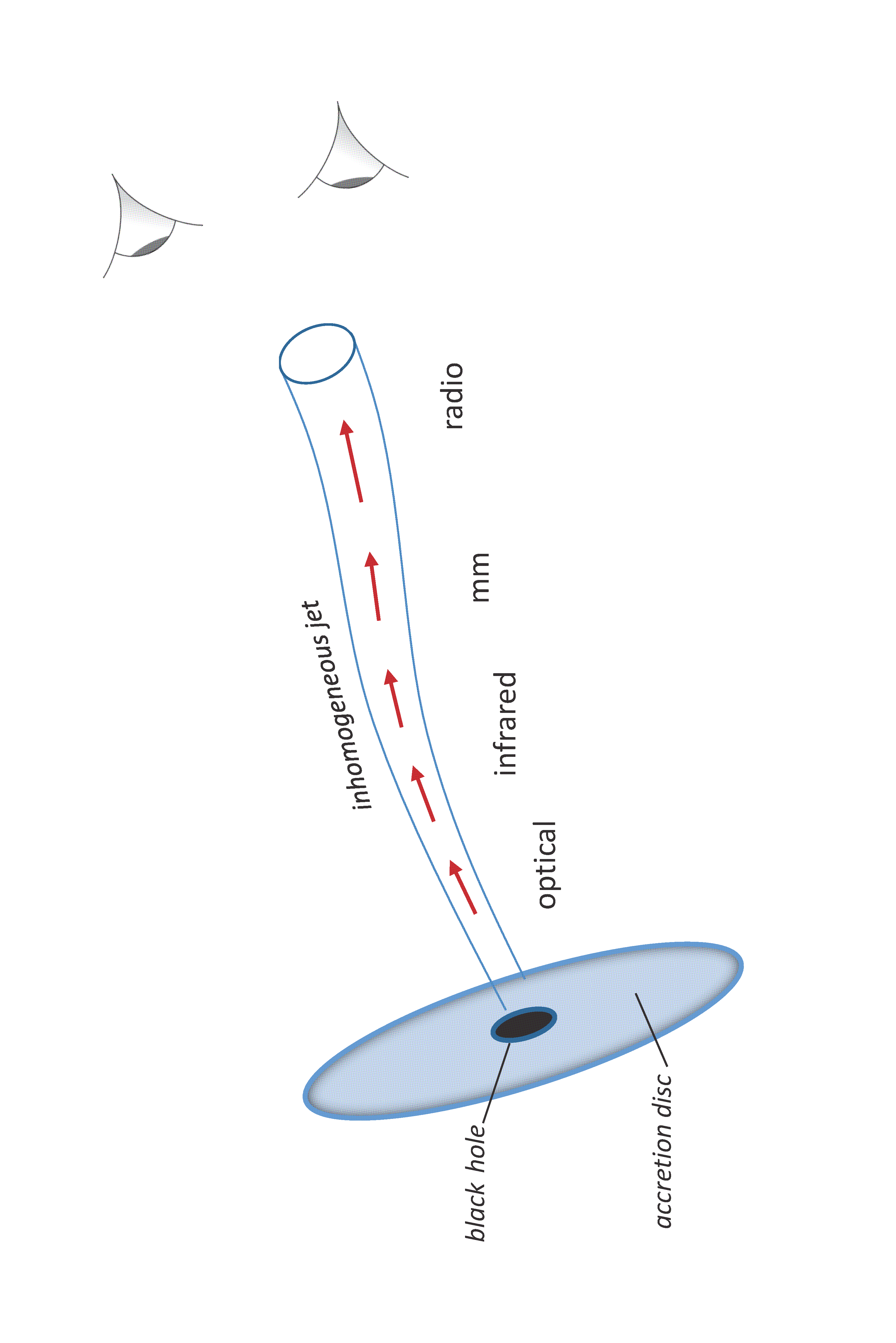}
\caption{{\bf A schematic representation of the proposed inhomogeneous jet model.} 
Photons of different frequencies come from diverse jet regions. Because of the jet curvature, these regions have different orientation. Therefore, the corresponding emission is more/less beamed depending on the better/worse alignment with the line of sight. The jet structure is dynamic, so that the orientation of each region changes in time. The two observing eyes on the right represent two different alignments of the line of sight relative to the jet. The ``upper" observer will see enhanced optical activity and relatively low mm--radio flux, while for the ``lower" observer the most beamed radiation is the mm one. }
\label{jet}
\end{figure}

The above scenario implies that the emission at different frequencies comes from different regions along a continuous jet (i.e., the jet is inhomogeneous), with different orientations with respect to the line of sight, and that the orientation is variable in time. 
A schematic representation of our model is given in Fig.\ \ref{jet}.

The variations in  $\theta$, $\delta$, and flux (Figs.\ \ref{sed} and \ref{radop}) have smaller amplitudes and are smoother in the radio and mm bands compared to those seen at shorter wavelengths. 
Within the model, this is most likely due to the fact that the radio and mm emitting regions are significantly more extended along the curved jet than the regions emitting the optical and near-IR light. Less dramatic variability would be expected from a larger emission region since the observations average the emission over a greater span of angles to the line of sight.

We tested the proposed geometrical model by comparing predicted and observed SEDs (Fig.\ \ref{sed}). For a given epoch, the predicted SED is obtained by summing the thermal emission model with a synchrotron SED that is derived by beaming the base-level SED with a frequency-dependent $\delta (\nu)$. Details are given in Methods. The agreement between model and data is very good.

We also analysed optical polarimetric data (see Methods and Extended Data Fig.\ \ref{pola}). 
The polarization percentage shows strong variability over all the considered period, but no general correlation with the flux, suggesting a mainly stochastic process due to turbulence$^{\ref{mar14}}$ or a variable jet direction$^{\ref{lyu17}}$. On the other side, the polarization angle undergoes wide rotations and in some cases its behaviour is consistent with the picture of a rotating twisted jet.

\bigskip\noindent
{\bf References}

\begin{enumerate}
\item{\label{bla79}
Blandford, R. D. \& K\"onigl, A. Relativistic jets as compact radio sources. {\it Astrophys. J.} {\bf 232}, 34-48 (1979)
}
\item{\label{ghi02}
Ghisellini, G., Celotti, A. \& Costamante, L. Low power BL Lacertae objects and the blazar sequence. Clues on the particle acceleration process. {\it Astron. Astrophys.} {\bf 386}, 833-842 (2002)
}
\item{\label{mar85}
Marscher, A. P. \& Gear, W.K. Models for high-frequency radio outbursts in extragalactic sources, with application to the early 1983 millimeter-to-infrared flare of 3C 273. {\it Astrophys. J.} {\bf 298}, 114-127 (1985)
}
\item{\label{sik01}
Sikora, M., B\l a\.zejowski, M., Begelman, Mitchell C. \& Moderski, R. Modeling the Production of Flares in Gamma-Ray Quasars. {\it Astrophys. J.} {\bf 554}, 1-11 (2001)
}
\item{\label{mar14}
Marscher, A. P. Turbulent, Extreme Multi-zone Model for Simulating Flux and Polarization Variability in Blazars. {\it Astrophys. J.} {\bf 780}, 87 (2014)
}
\item{\label{vil99}
Villata, M. \& Raiteri, C.M. Helical jets in blazars I. The case of Mkn 501. {\it Astron. Astrophys.} {\bf 347}, 30-36 (1999)
}
\item{\label{mar08}
Marscher, A. P. et al. The inner jet of an active galactic nucleus as revealed by a radio-to-gamma-ray outburst. {\it Nature} {\bf 452}, 966-969 (2008)
}
\item{\label{abd10}
Abdo, A. A. et al. A change in the optical polarization associated with a $\gamma$-ray flare in the blazar 3C~279. {\it Nature} {\bf 463}, 919-923 (2010)
}
\item{\label{lar16}
Larionov, V. M. et al. Exceptional outburst of the blazar CTA 102 in 2012: the GASP-WEBT campaign and its extension {\it Mon. Not. R. Astron. Soc.} {\bf 461}, 3047-3056 (2016)
}

\item{\label{cas15}
Casadio, C. et al. A Multi-wavelength polarimetric study of the blazar CTA 102 during a gamma-ray flare in 2012. {\it Astrophys. J.} {\bf 813}, 51-64 (2015)
}

\item{\label{cam92}
Camenzind, M. \& Krockenberger M. The lighthouse effect of relativistic jets in blazars - A geometric origin of intraday variability. {\it Astron. Astrophys.} {\bf 255}, 59-62 (1992)
}
\item{\label{ost04}
Ostorero, L., Villata, M. \& Raiteri, C. M. Helical jets in blazars Interpretation of the multifrequency long-term variability of AO 0235+16. {\it Astron. Astrophys.} {\bf 419}, 913-925 (2004)
}

\item{\label{rie04}
Rieger, F. M. On the geometrical origin of periodicity in blazar-type sources. {\it Astrophys. J.} {\bf 615}, L5-L8 (2004)
}

\item{\label{vil09}
Villata, M. et al. The correlated optical and radio variability of BL Lacertae. WEBT data analysis 1994-2005. {\it Astron. Astrophys.} {\bf 501}, 455-460 (2009)
}
\item{\label{ghi09}
Ghisellini G. \& Tavecchio, F. Canonical high-power blazars. {\it Mon. Not. R. Astron. Soc.} {\bf 397}, 985-1002 (2009)
}
\item{\label{mar17}
Marcotulli, L. et al. High-redshift blazars through {\it NuSTAR} eyes. {\it Astrophys. J.} {\bf 839}, 96 (2017)
}
\item{\label{mig10}
Mignone, A., Rossi, P., Bodo, G., Ferrari, A. \& Massaglia, S. High-resolution 3D relativistic MHD simulations of jets. {\it Mon. Not. R. Astron. Soc.} {\bf 402}, 7-12 (2010)
}
\item{\label{vil06}
Villata, M. et al. The unprecedented optical outburst of the quasar 3C 454.3. The WEBT campaign of 2004-2005. {\it Astron. Astrophys.} {\bf 453}, 817-822 (2006)
}
\item{\label{rai14}
Raiteri, C. M. et al. Infrared properties of blazars: putting the GASP-WEBT sources into context. {\it Mon. Not. R. Astron. Soc.} {\bf 442}, 629-646 (2014)
}
\item{\label{imp88}
Impey, C. D. \& Neugebauer, G. Energy distributions of blazars. {\it Astron. J.} {\bf 95}, 307-351 (1988)
}
\item{\label{mal11}
Malmrose, M. P.,  Marscher, A. P., Jorstad, S. G., Nikutta, R. \& Elitzur, M. Emission from hot dust in the infrared spectra of gamma-ray bright blazars. {\it Astrophys. J.} {\bf 732}, 116 (2011)
}
\item{\label{urr95}
Urry, C. M. \& Padovani, P. Unified schemes for radio-loud Active Galactic Nuclei. {\it Publ. Astron. Soc. of the Pacific} {\bf 107}, 803-845, (1995)
}
\item{\label{sav10}
Savolainen, T. et al. Relativistic beaming and gamma-ray brightness of blazars. {\it Astron. Astrophys.} {\bf 512}, A24 (2010)
}
\item{\label{bri17}
Britzen, S. et al. A swirling jet in the quasar 1308+326. {\it Astron. Astrophys.} {\bf 602}, A29 (2017)
}
\item{\label{per12}
Perucho, M., Kovalev, Y. Y., Lobanov, A. P., Hardee, P. E. \& Agudo, I. Anatomy of helical extragalactic jets: the case of S5 0836+710. {\it Astrophys. J.} {\bf 749}, 55 (2012)
}
\item{\label{fro13}
Fromm, C. M. et al. Catching the radio flare in CTA 102. III. Core-shift and spectral analysis. {\it Astron. Astrophys.} {\bf 557}, A105 (2013)
}

\item{\label{lyu17}
Lyutikov, M. \& Kravchenko, E. V. Polarization swings in blazars. {\it Mon. Not. R. Astron. Soc.} {\bf 467}, 3876 (2017)
}
\end{enumerate}

\bigskip\noindent
{\bf Author Contributions}
C.M.R. and M.V. managed the WEBT observing campaign, analysed the data, developed the geometrical interpretation and wrote the manuscript. 
J.A.A-P, A.A.A., M.I.C., N.C.-S., N.V.E., A.D.P., A.G., C.L., F.P., C.P., F.J.R.-L., G.R.-C. performed near-infrared and optical observations and related data reduction.
I.A., C.C., A.F., J.L.G., S.N.M. performed photometric and polarimetric optical and radio observations and related data reduction.
E.B., J.E.,  C.E., T.S.G., D.H., S.G.J., M.J., E.N.K., V.M.L., E.G.L., L.V.L., M.P.M., A.P.M., R.M., A.A.M., J.W.M., D.A.M., S.S.S., Yu.V.T., I.S.T., A.A.V. acquired and reduced optical photometric and polarimetric data.
R.B., G.V.B., G.A.B., V.B., M.S.B., P.C., D.C., W.-P.C., G.D., Sh.A.E., H.J., B.J., K.K., O.M.K., S.O.K., C.S.L., K.M., B.McB., B.M., M.M., D.O.M., S.V.N., M.G.N., J.M.O., D.N.O., E.O., T.A.P., N.R., K.S., A.C.S., M.R.S., E.S., B.A.S., L.S.-M., I.A.S., A.S., O.V. carried out optical observations and related data reduction.
M.A.G., A.L., J.T., C.T., M.T. performed radio observations and related data reduction.
W.B. acquired and reduced optical spectra.
T.P. made optical photometric and spectroscopic observations and related data reduction.
P.S.S. carried out optical photometric, polarimetric, and spectroscopic observations and reduced the data.
F.D. and all the above authors reviewed and contributed to the manuscript.

\bigskip\noindent
{\bf Author information}
Reprints and permissions information is available at www.nature.com/reprints. The authors declare no competing financial interests. Correspondence and requests for materials should be addressed to C.M.R. (Email: raiteri@oato.inaf.it) or M.V. (Email: villata@oato.inaf.it)

\bigskip\noindent
{\bf Methods}

\noindent
{\bf Observations.}

Born in 1997, the Whole Earth Blazar Telescope (WEBT) is an international collaboration of astronomers monitoring blazars in the optical, near-infrared, mm and radio bands to investigate these highly active objects. 
Optical data for this paper were acquired at the following observatories: 
Abastumani (Georgia),
AstroCamp (Spain),
Belogradchik (Bulgaria),
Calar Alto (Spain),
Campo Imperatore (Italy),
Crimean (Russia),
Kitt Peak (USA),
Lowell (USA; 70 cm, DCT and Perkins telescopes),
Lulin (Taiwan),
Michael Adrian (Germany),
Mt.\ Maidanak (Uzbekistan),
New Mexico Skies (USA),
Osaka Kyoiku (Japan),
Polakis (USA),
Roque de los Muchachos (Spain; Liverpool, NOT and TNG telescopes),
ROVOR (USA),
Rozhen (Bulgaria; 200 and 50/70 cm telescopes),
San Pedro Martir (Mexico),
Sirio (Italy),
Skinakas (Greece),
Steward (USA; Kuiper, Bok, and Super-LOTIS),
St.\ Petersburg (Russia),
Teide (Spain),
Tien Shan (Kazakhstan),
Tijarafe (Spain),
Tucson (USA),
Valle d'Aosta (Italy),
Vidojevica (Serbia)
within a WEBT project.
The source magnitude was calibrated using common photometric standard stars in the source field (Star 1 and Star 2 by [\ref{rai98}], with the addition of other stars from [\ref{dor13}], when needed). This minimises possible offsets among different data sets. 
Observations were performed in the Johnson-Cousins' $BVRI$ bands, except for those at the NOT and Liverpool telescopes. The NOT data were provided in the Sloan Digital Sky Survey $ugriz$ filters and then converted with the transformations by [\ref{jor06}]. 
The Liverpool data were taken with the ``Red" (770--1000 nm), ``Green" (650--760 nm) and ``Blue" (350--640 nm) cameras of the RINGO3 instrument; they were transformed to the nearest conventional Johnson-Cousins bands via shifts derived from periods of overlapping data with other instruments.

Near-infrared data were taken in the $JHK$ filters at the Campo Imperatore, Lowell (Perkins) and Teide observatories. Data reduction is described in [\ref{rai14}]. 

Extended Data Fig.\ \ref{webt} shows the optical and near-infrared light curves of CTA~102 in the last four observing seasons.
During the most dramatic phases of the 2016--2017 outburst some episodes of noticeable and well-sampled intranight variability have been observed. Four of them are shown in Extended Data Fig.\ \ref{idv}.

Observations in the radio and mm bands were performed with the 14-m radio telescope of the Mets\"ahovi Radio Observatory (37 GHz) in Finland, by the 30-m IRAM telescope\ (86 and 230 GHz) in Spain, and by the Submillimeter Array (230 GHz) in Hawaii, USA. For details of the radio data analysis see [\ref{ter98}, \ref{agu10}, \ref{gur07}].

\bigskip\noindent
{\bf Spectral behaviour.}

The optical spectral behaviour of CTA 102 in the same period of Extended Data Fig.\ \ref{webt} is shown in Extended Data Fig.\ \ref{colori}, where $B-R$ colour indices (and spectral slopes $\alpha$ of the $F \propto \nu^{-\alpha}$ law) are obtained by coupling data taken by the same telescope within 15 minutes.
A redder-when-brighter trend until $R \sim 15$ (Spearman's rank correlation coefficient $\rho=-0.82$) is followed by a slight bluer-when-brighter trend ($\rho=0.26$) as the source flux increases.
This has previously been noticed$^{\ref{vil06}}$ for 3C~454.3 and means that as the source brightens the disc contribution becomes negligible and then the spectrum becomes bluer again, possibly because of changes in the Doppler factor$^{\ref{lar10}}$.
In particular, we note that in faint states, when the source is ``blue", the $B$-band brightness is much more stable than that in the $R$-band, which means that the $B$-band flux is dominated by the disc emission, but the $R$-band flux still receives important synchrotron contribution from the jet, which makes the colour index significantly vary also in these states.

Spectroscopic observations in the optical band were carried out at the Steward (Kuiper, Bok, and MMT telescopes) and Roque de los Muchachos (TNG and NOT telescopes) observatories. 
A selection of these spectra is shown in Extended Data Fig.\ \ref{spettri}. Those taken during faint states show in particular a prominent, broad Mg~II emission feature and hard spectral shape; the lines gradually disappear and the spectra soften as the brightness rises, as a consequence of the increasing importance of the featureless synchrotron continuum over the BBB.
At the highest flux levels the optical spectra harden again, consistent with the bluer-when-brighter behaviour noticed above.

\bigskip\noindent
{\bf Modelling the thermal emission component.}

To model the BBB, whose contribution is assumed to remain constant throughout the period of study, we examine the flux variability ranges in the monitored bands.
In Fig.\ \ref{sed} we plot all data acquired in the 2008--2017 period by the WEBT observers in the radio--mm (37, 86 and 230 GHz), in the near-infrared ($KHJ$) and in the optical ($IRVB$) bands. 
When passing from observed magnitudes to flux densities we corrected for Galactic extinction using the prescriptions of the NASA/IPAC Extragalactic Database (NED). We build the SED of a putative synchrotron minimum brightness state by fitting a log-parabolic model through the observed radio--mm minimum flux densities and a minimum synchrotron flux density in $K$ band, which is obtained by assuming that the observed minimum flux density in that band receives equal synchrotron and thermal contributions. The adequacy of a log-parabolic model to describe the broad-band synchrotron emission of blazars has been discussed by e.g.\ [\ref{mas04}] and this model is widely used. The thermal contribution from the BBB in all near-infrared and optical bands is then obtained by subtracting the model minimum synchrotron flux from the observed flux minima. The result is in agreement with that derived for the same object by [\ref{rai14}] using a QSO template. 

To complete the AGN model toward the mid--far infrared, we also added the emission contribution from the dust torus as obtained by [\ref{rai14}], even if it is always negligible, except for the case when the source is very faint.

\bigskip\noindent
{\bf Relativistic beaming theory}

The Doppler factor is defined as  $\delta=[\Gamma (1-\beta \cos \theta)]^{-1}$, where $\beta$ is the emitting source bulk velocity in units of the speed of light, $\Gamma= (1-\beta^2)^{-1/2}$ is the corresponding Lorentz factor, and $\theta$ is the viewing angle. 
Any time interval is shortened in the observer's frame as $\Delta t = \Delta t' / \delta$, while frequencies are blue-shifted as $\nu=\delta \nu'$ (primed quantities refer to the source rest frame).

For a continuous jet with isotropic emission in the rest frame, the observed and rest-frame flux densities are linked by $F_\nu (\nu)=\delta^{2+\alpha} \, F'_{\nu '} (\nu)$, 
with $F'_{\nu '} \propto (\nu ')^{-\alpha}$ [\ref{urr95}]. 

As a consequence, for a given beaming state characterized by a Doppler factor $\delta$, the flux variability amplitude due to possible intrinsic processes $\Delta F_\nu \propto \delta^{2+\alpha}$.
Therefore, a stronger beaming enhances not only the flux, but also the variability amplitude of the intrinsic variations (fast flares), beside shortening the variability time scale. 

We can correct the observed flux densities at a given frequency for the variable Doppler beaming effect according to:
$F_\nu^{\rm cor}= F_\nu \, (\delta_\nu^{\rm const} / \delta_\nu^{\rm var})^{2+\alpha_\nu}$,
where $F_\nu^{\rm cor}$  represent the values that we would observe in case the jet had a constant orientation resulting in $\delta_\nu^{\rm const}$. In Fig.\ \ref{radop} we present $F_\nu^{\rm cor}$ in the $R$ band under the choice $\delta_\nu^{\rm const}=\delta_{\rm base}$.

\bigskip\noindent
{\bf Variability time scales.}

Inspection of the optical light curves reveals that variability time scales shorten when the source is brighter, which is a robust indication that the long-term flux changes are due to Doppler factor variations. 
This can be verified quantitatively by performing time series analysis, i.e.\ investigating the time structure of flux variations. In order to avoid possible bias due to the long-term trend and related difference in flux amplitude, we applied the analysis to the flux densities corrected for the variable relativistic beaming effect (see Fig.\ \ref{radop}).
We separated the dataset in two subsamples, corresponding to bright ($\delta > \delta_{\rm max}/2$) and faint ($\delta < \delta_{\rm max}/2$) states, thus essentially separating the huge 2016--2017 outburst from the rest of the data. For the two subsamples we calculated the structure function$^{\ref{sim85}}$ (SF), which 
measures the square mean difference in the flux densities as a function of the time separation $\tau$.
The results are shown in Extended Data Fig.\ \ref{tisca}, where the minimum variability time scale corresponds to the SF first peak, which is about 4 days for the bright states, and about 8 days for the faint ones. This doubling of the time scales matches very well the $\delta$-halving criterion adopted to separate the two subsamples ($\Delta t \propto \delta^{-1}$).
We checked the SF results by means of discrete autocorrelation function$^{\ref{huf92}}$ (ACF), which is shown in the same figure. Here time scales are given by ACF minima. The shortest time scales for the high (low) flux states are confirmed to be about 4 (8) days.

We finally applied the Kolmogorov-Smirnov (K-S) statistic to check whether the SFs and ACFs of the two subsamples are drawn from the same distribution. The values of the K-S statistic are 0.67 for the SFs and 0.39 for the ACFs. Their significance levels are $1.1 \times 10^{-6}$ for the SFs and $1.0 \times 10^{-4}$ for the ACFs; such small values mean that the two distributions are significantly different.

In view of these results and in agreement with the relativistic beaming theory, to model the optical long-term trend we set an adaptive bin size that reduces the time bin by a factor $n$ as the flux increases by a factor $n^{2+\alpha}$, with $n=2,3,4,5,6,7$ and $\alpha=1.7$, the slope of the minimum synchrotron spectrum in the $R$ band.

\bigskip\noindent
{\bf Assumption on the Lorentz factor.}

We have assumed a single $\Gamma$ value throughout the jet region of interest. Other scenarios, with $\Gamma$ varying along, or transversally to, the jet are in principle possible and have been adopted in some cases. 

However, the portion of the jet we are considering, i.e.\ that emitting the bulk of photons from the optical band down to 37 GHz, is inside the very inner core of the radio images, so that we can easily guess that the Lorentz factor does not change significantly along this region. On the other side, transverse velocity gradients, for which there is no observing evidence for the inner zones of blazar jets, are sometimes introduced to explain the properties of high-energy, Comptonized radiation$^{\ref{ghi05},\ref{sik16}}$, but they would represent an unnecessary complication for the purposes of our model.

\bigskip
\bigskip\noindent
{\bf Comparison between predicted and observed SEDs.}

The frequency-dependent trend of the Doppler factor necessary to obtain a model SED for a given epoch is derived by interpolating/extrapolating the values of $\delta$ at 37 GHz, 230 GHz and in the $R$ band at that epoch (Fig.\ \ref{radop}) in the $\delta$ versus $\log \nu$ space, after correcting frequencies for the different beaming affecting the base-level SED and the model SED for the considered epoch. 
We perform linear interpolation of the Doppler factor in the $\delta$ versus $\log \nu$ space by minimizing the chi-square error statistic. In case the unreduced chi-square goodness-of-fit statistic was greater than 1.2, indicating poor fit, we also performed a parabolic interpolation and took the average fit between the two. This occurred three times, for the epochs $\rm JD-2,450,000=7637$, 7654, 7691.

In Fig.\ \ref{sed} ten SEDs are displayed, corresponding to selected epochs spanning the source variability range. The predicted and observed SEDs are in very good agreement. In particular, the spectral slope of the optical part of the models matches very well those of the observed optical spectra. 
The offset between the optical and near-infrared data at epoch $\rm JD-2,450,000=7717$ is likely due to non perfect simultaneity of the observations coupled with strong intranight variability.
In the figure we also show the corresponding viewing angle as a function of frequency for all epochs. The brightest state during the 2016--2017 optical outburst corresponds to the maximum difference of orientation (maximum misalignment) between the radio and optical emitting zones when the optical zone has the best alignment with the line of sight.

\bigskip\noindent
{\bf Polarization.}

Optical polarimetric data were acquired by seven observatories: Calar Alto, Crimean, Lowell (Perkins), ROVOR, San Pedro Martir, Steward, St.\ Petersburg. The temporal behaviour of the jet polarization percentage $P_{\rm jet}$ and electric vector polarization angle (EVPA) are shown in Extended Data Fig.\ \ref{pola}. 

$P_{\rm jet}$ was obtained from the observed polarization degree $P$ by correcting for the dilution effect of the BBB unpolarized emission:
$P_{\rm jet}=P \times F/F_{\rm jet}$,
where $F$ is the de-absorbed flux density, and $F_{\rm jet}$ is the same quantity after subtraction of the BBB flux contribution.
In the figure we also plot the mean value of $P_{\rm jet}$ for the whole period and the mean values and standard deviations for each observing season. 
The EVPA was adjusted for the $\pm n \times \pi$ ($n \in N$) ambiguity by requiring that the angle difference between subsequent points within the same observing season is minimum.  

Strong variability of $P_{\rm jet}$ and wide rotations of EVPA both in the clockwise and anticlockwise direction are observed over all the considered period. 
No general correlation is found with the observed flux or with the flux variations that are left after correction for the variable Doppler beaming (see Fig.\ \ref{radop}).
The only hint of correlation between polarization and flux variations is given by the concomitance of minima in the viewing angle (flux peaks) with either fast rotation of EVPA (at $\rm JD \sim 2,457,300$) or inversion of its direction of rotation (at $\rm JD \sim 2,456,200$  and  $\rm JD \sim 2,457,750$, i.e.\ during the peaks of the 2012$^{\ref{lar16},\ref{cas15}}$ and 2016--2017 outbursts).
Both situations may occur when considering a rotating helical jet with longitudinal magnetic field. As the helix rotates and the emitting region approaches the line of sight, the EVPA undergoes a fast rotation or a change of the rotation direction  depending on whether the angle between the line of sight and the helix axis is smaller or larger than the helix pitch angle, respectively.
However, some turbulence must be present$^{\ref{mar14}}$ to explain the scrambled behaviour of $P_{\rm jet}$. 
Alternatively, [\ref{lyu17}] showed that apparent random behaviour of $P$ (and $F$) can accompany large EVPA swings in a jet with helical magnetic field and variable propagation direction.

\bigskip\noindent
{\bf Comparison with a standard one-zone model.}

We have shown that the CTA~102 long-term multiwavelength variability is well explained by changes of the Doppler factor. We now investigate whether also commonly used one-zone models can explain the source spectral changes in this way.
In Extended Data Fig.\ \ref{one} we present the results obtained with the standard one-zone model developed by [\ref{tra09}]. We started to fit the SED at $\rm JD=2,457,637$, which represents an intermediate flux level. The physical parameters used are: 	
blob radius $\log R=17.8$ [cm]; magnetic field $B=0.08 \, \rm G$; Doppler factor $\delta=21.5$; number of emitting electrons $N=30 \, \rm cm^{-3}$; electron energies between $\log \gamma_{\rm min}=1$ and $\log \gamma_{\rm max}=5$ and a power-law+cut-off electron energy distribution with $\alpha=2.15$ and $\log \gamma_{\rm cut}=3.6$.

We then tried to fit the highest and the lowest optical levels shown in Fig.\ \ref{sed} by changing only $\delta$, which from our results is the essential parameter to be changed. The brightest state requires $\delta=40$ and the faintest $\delta=9.5$. However, the model fits do not match the lower-frequency data; in particular, the flux in the mm band is largely over- or under-produced. 
Of course, better fits could be obtained with the one-zone model, but at the cost of changing a number of parameters, in particular the electron energy distribution.
One should then check if a reasonable temporal evolution of all these parameters can be found to explain the multiwavelength light curves.

\bigskip\noindent
{\bf Methods references}

\begin{enumerate}
\setcounter{enumi}{27}
\item{\label{rai98}
Raiteri C.M., Villata, M., Lanteri, L., Cavallone, M. \& Sobrito, G. $BVR$ photometry of comparison stars in selected blazar fields. II. Photometric sequences for 9 quasars. {\it Astron. Astrophys. Suppl. Ser.} {\bf 130}, 495-500 (1998)
}

\item{\label{dor13}
Doroshenko, V. T. et al. BVRI CCD-Photometry of Comparison Stars in the Fields of
Galaxies with Active Nuclei. V. {\it Astrophysics} {\bf 56}, 343-358 (2013)
}

\item{\label{jor06}
Jordi, K., Grebel, E. K. \& Ammon, K. Empirical color transformations between SDSS photometry and other photometric systems. {\it Astron. Astrophys.} {\bf 460}, 339-347 (2006)
}

\item{\label{ter98}
Ter\"asranta, H. et al. Fifteen years monitoring of extragalctic radio sources at 22, 37 and 87 GHz. {\it Astron. Astrophys. Suppl. Ser.} {\bf 132}, 305-331 (1998)
}

\item{\label{agu10}
Agudo, I, Thum, C., Wiesemeyer, H. \& Krichbaum, T. P. A 3.5 mm Polarimetric Survey of Radio-loud Active Galactic Nuclei. {\it Astrophys. J. Suppl. Ser.} {\bf 189}, 1-14 (2010)
}

\item{\label{gur07}
Gurwell, M. A., Peck, A. B., Hostler, S. R., Darrah, M. R. \& Katz, C. A. Monitoring Phase Calibrators at Submillimeter Wavelengths. in {\it From Z-Machines to ALMA: (Sub)Millimeter Spectroscopy of Galaxies (ASP Conf. Ser. 375)} (eds Baker, A. J., Glenn, J., Harris, A. I., Mangum, J. G. \& Yun, M. S.) 234-237 (Astronomical Society of the Pacific, San Francisco, 2007)
}

\item{\label{lar10}
Larionov, V. M., Villata, M. \& Raiteri, C.M. The nature of optical and near-infrared variability of BL Lacertae. {\it Astron. Astrophys.} {\bf 510}, A93 (2010)
}

\item{\label{mas04}
Massaro, E., Perri, M., Giommi, P. \& Nesci, R. Log-parabolic spectra and particle acceleration in the BL Lac object Mkn 421: Spectral analysis of the complete BeppoSAX wide band X-ray data set.
{\it Astron. Astrophys.} {\bf 413}, 489-503 (2004)
}

\item{\label{sim85}
Simonetti, J. H., Cordes, J. M. \& Heeschen, D. S. Flicker of extragalactic radio sources at two frequencies. {\it Astrophys. J.} {\bf 296}, 46--59 (1985)
}

\item{\label{huf92}
Hufnagel, B. R. \& Bregman, J. N. Optical and radio variability in blazars. {\it Astrophys. J.} {\bf 386}, 473--484 (1992)
}

\item{\label{ghi05}
Ghisellini, G., Tavecchio, F. \& Chiaberge, M. Structured jets in TeV BL Lac objects and radiogalaxies Implications for the observed properties. {\it Astron. Astrophys.} {\bf 432}, 401-410 (2005)
}

\item{\label{sik16}
Sikora, M., Rutkowski, M. \& Begelman, M. C. A spine-sheath model for strong-line blazars. {\it Mon. Not. R. Astron. Soc.} {\bf 457}, 1352-1358 (2016)
}


\item{\label{tra09}
Tramacere, A., Giommi, P., Perri, M., Verrecchia, F. \& Tosti, G. {\it Swift} observations of the very intense flaring activity of Mrk 421 during 2006. I. Phenomenological picture of electron acceleration and predictions for MeV/GeV emission. {\it Astron. Astrophys.} {\bf 501}, 879-898 (2009)
}

\end{enumerate}

\newpage\noindent
{\bf Acknowledgements}

\noindent
Data from the Steward Observatory spectropolarimetric monitoring project were used.  This program is supported by NASA/Fermi Guest Investigator grant NNX15AU81G.
We thank Dagmara Oszkiewicz, Jyri Lehtinen, Fatemeh Sadat Kiaeerad, 
Stefan Geier, Ren\'e Tronsgaard Rasmussen, Pere Blay Serrano
and NEON summer school students Sonia Tamburri, Benjamin Hendricks,
Marijana Smailagic, Jaan Laur for assistance with the NOT 
observations and Albino Carbognani for assistance with the OAVdA 
observations.
Based on observations made with the Nordic Optical Telescope, operated by 
the Nordic Optical Telescope Scientific Association at the Observatorio 
del Roque de los Muchachos, La Palma, Spain, of the Instituto de 
Astrofisica de Canarias.
This article is partly based on observations made with the telescopes IAC80 and TCS operated by the Instituto de Astrofisica de Canarias in the Spanish Observatorio del Teide on the island of Tenerife. The IAC team acknowledges the support from the group of support astronomers and telescope operators of the Observatorio del Teide. Based (partly) on data obtained with the STELLA robotic telescopes in Tenerife, an AIP facility jointly operated by AIP and IAC.
The Abastumani team acknowledges financial support by the Shota Rustaveli
NSF under contract FR/217950/16. Kurtanidze O. M. acknowledges the financial support
of NSF of China under grant U1531245.
Kazakhstan team acknowledges financial support by the Ministry
of Education and Science under grants No.0073-7/PSF and No.0263/PSF.
This research was partially supported by the Bulgarian National Science
Fund of the Ministry of Education and Science under grant DN 08-1/2016.
The Skinakas Observatory is a collaborative project of the University of
Crete, the Foundation for Research and Technology -- Hellas, and the
Max-Planck-Institut f\"ur Extraterrestrische Physik.
The BU group acknowledges a support from National Science Foundation grant AST-1615796. 
G.Damljanovic and O.Vince gratefully acknowledge the
observing grant support from the Institute of Astronomy and Rozhen National
Astronomical Observatory, Bulgaria Academy of Sciences, via bilateral joint
research project "Observations of ICRF radio-sources visible in optical domain"
(the head is Dr. G.Damljanovic). This work is a part of the Projects No 176011
(Dynamics and kinematics of celestial bodies and systems), No 176004
(Stellar physics) and No 176021 (Visible and invisible matter in nearby
galaxies: theory and observations) supported by the Ministry of Education,
Science and Technological Development of the Republic of Serbia.
This research was partially supported by the Bulgarian National Science
Fund of the Ministry of Education and Science under grant DN 08-1/2016.
St.Petersburg University team acknowledges support from Russian Science Foundation grant 17-12-01029.
AZT-24 observations are made within an agreement between  Pulkovo,
Rome and Teramo observatories.
The research  at Ulugh Beg Astronomical Institute was supported by grants N. BA-FA-F-2-007  of the Uzbekistan Agency for Science and Technology.
Based (partially) upon observations carried out at the Observatorio Astron\'omico Nacional on the Sierra San Pedro M\'artir (OAN-SPM), Baja California, M\'exico.
The Astronomical Observatory of the Autonomous Region of the Aosta Valley (OAVdA) is managed by the Fondazione Cl\'ement Fillietroz-ONLUS, which is supported by the Regional Government of the Aosta Valley, the Town Municipality of Nus and the ``Unit\'e des Communes vald\^{o}taines Mont-\'Emilius".
This paper is partly based on observations carried out with the IRAM 30~m, the Calar Alto 2.2~m, and the Liverpool 2.0~m Telescopes. 
Calar Alto data was acquired as part of the MAPCAT project (www.iaa.es/$\sim$iagudo/\textunderscore iagudo/MAPCAT.html). IRAM data was acquired as part of the POLAMI project (polami.iaa.es).
IRAM is supported by INSU/CNRS (France), MPG (Germany) and IGN (Spain). Calar Alto observatory is jointly operated by MPIA (Germany) and the IAA-CSIC (Spain). The Liverpool Telescope is operated by JMU with financial support from the UK-STFC. IA acknowledges support by a Ram\'on y Cajal grant of the Ministerio de Econom\'ia y Competitividad (MINECO) of Spain. The research at the IAA--CSIC was supported in part by the MINECO through grants AYA2016--80889--P, AYA2013--40825--P, and AYA2010--14844, and by the regional government of Andaluc\'{i}a through grant P09--FQM--4784.
This publication makes use of data obtained at the Mets\"ahovi Radio Observatory, operated by the Aalto University.
The Submillimeter Array is a joint project between the Smithsonian Astrophysical Observatory and the Academia Sinica Institute of Astronomy and Astrophysics and is funded by the Smithsonian Institution and the Academia Sinica.
This research has made use of the NASA/IPAC Extragalactic Database (NED) which is operated by the Jet Propulsion Laboratory, California Institute of Technology, under contract with the National Aeronautics and Space Administration. 
This research has made use of NASA's Astrophysics Data System.
Part of this work is based on archival data, software or online services provided by the ASI Science Data Center (ASDC).

\bigskip\noindent
{\bf Data Availability}
Data taken and assembled by the WEBT collaboration (optical, near-infrared and radio light curves) are stored in the WEBT archive at the Osservatorio Astrofisico di Torino - INAF 

\noindent
(http://www.oato.inaf.it/blazars/webt/); they become publicly available one year after publication and can be requested by contacting the WEBT President Massimo Villata (villata@oato.inaf.it).
Optical spectropolarimetric data from the Steward Observatory are public and can be downloaded from http://james.as.arizona.edu/$\sim$psmith/Fermi/

\bigskip\noindent
{\bf Extended Data}
\setcounter{figure}{0} 
   
\begin{figure}[h]
\center\includegraphics[width=11cm]{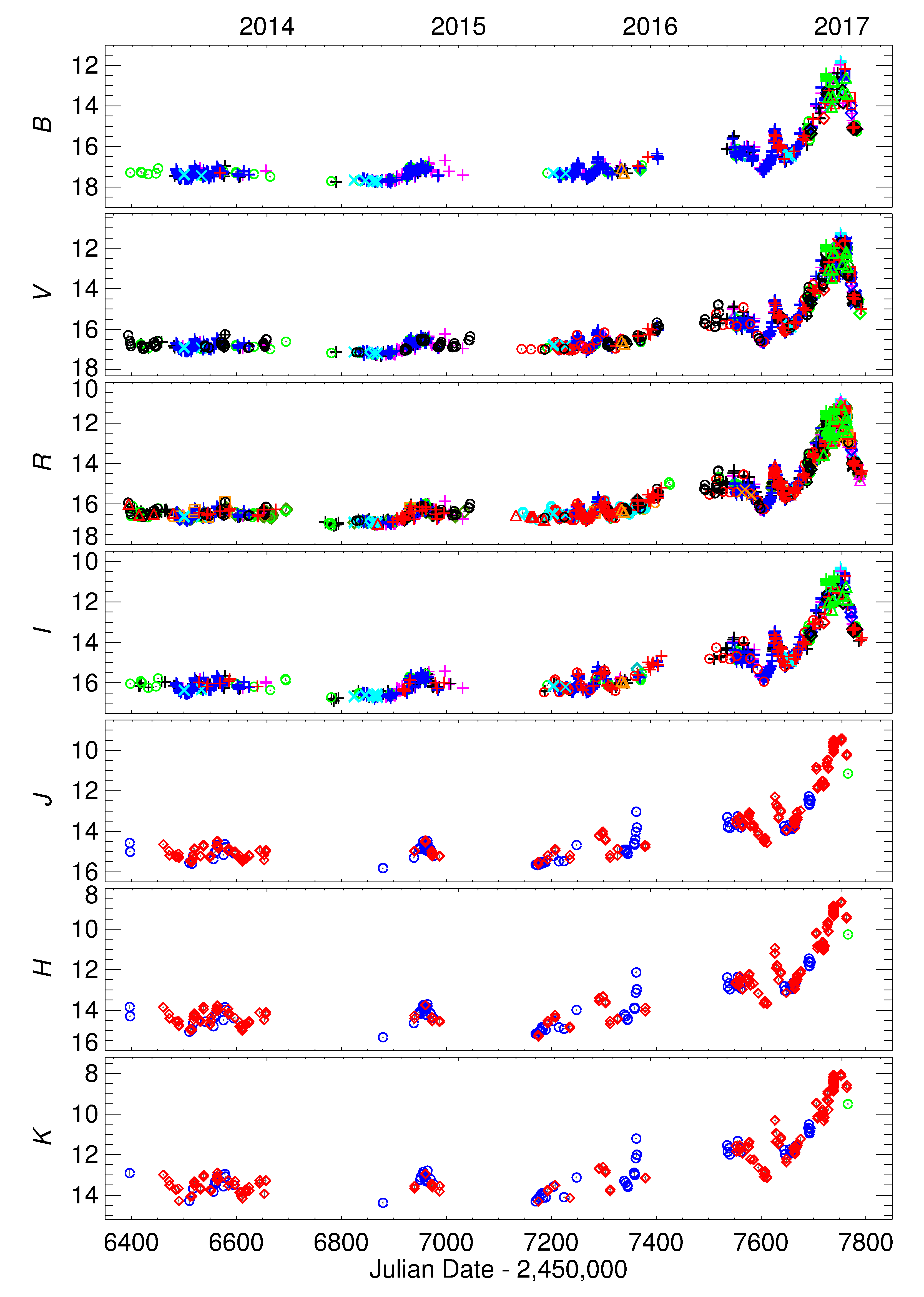}
\caption{{\bf Observed light curves of CTA~102 in the optical $BVRI$ and near-infrared $JHK$ bands.} They are built with data from 39 telescopes (marked with different symbols and colours) in 28 observatories participating to the WEBT project. Measure errors (1 s.d.) are smaller than the symbols size.}
\label{webt}
\end{figure}

\begin{figure}[h]
\center\includegraphics[width=\linewidth]{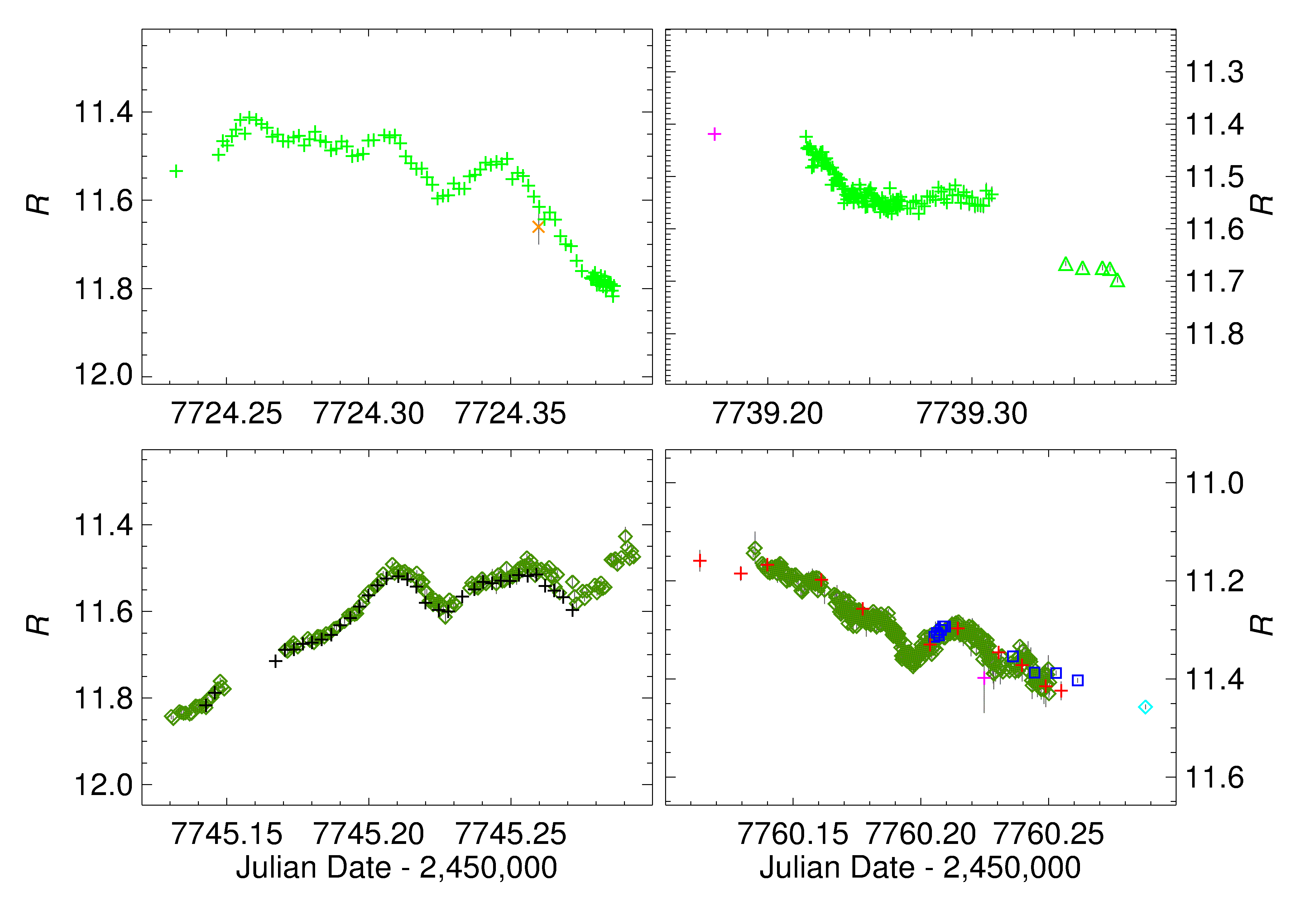}
\caption{{\bf Four episodes of noticeable and well-sampled intranight variability.} Enlargements of the $R$-band light curve of Fig.\ \ref{outburst} during the most dramatic phases of the 2016--2017 optical outburst reveal very fast brightness changes. Error bars represent 1 s.d. measure errors.}
\label{idv}
\end{figure}

\begin{figure}[h]
\center\includegraphics[width=12cm]{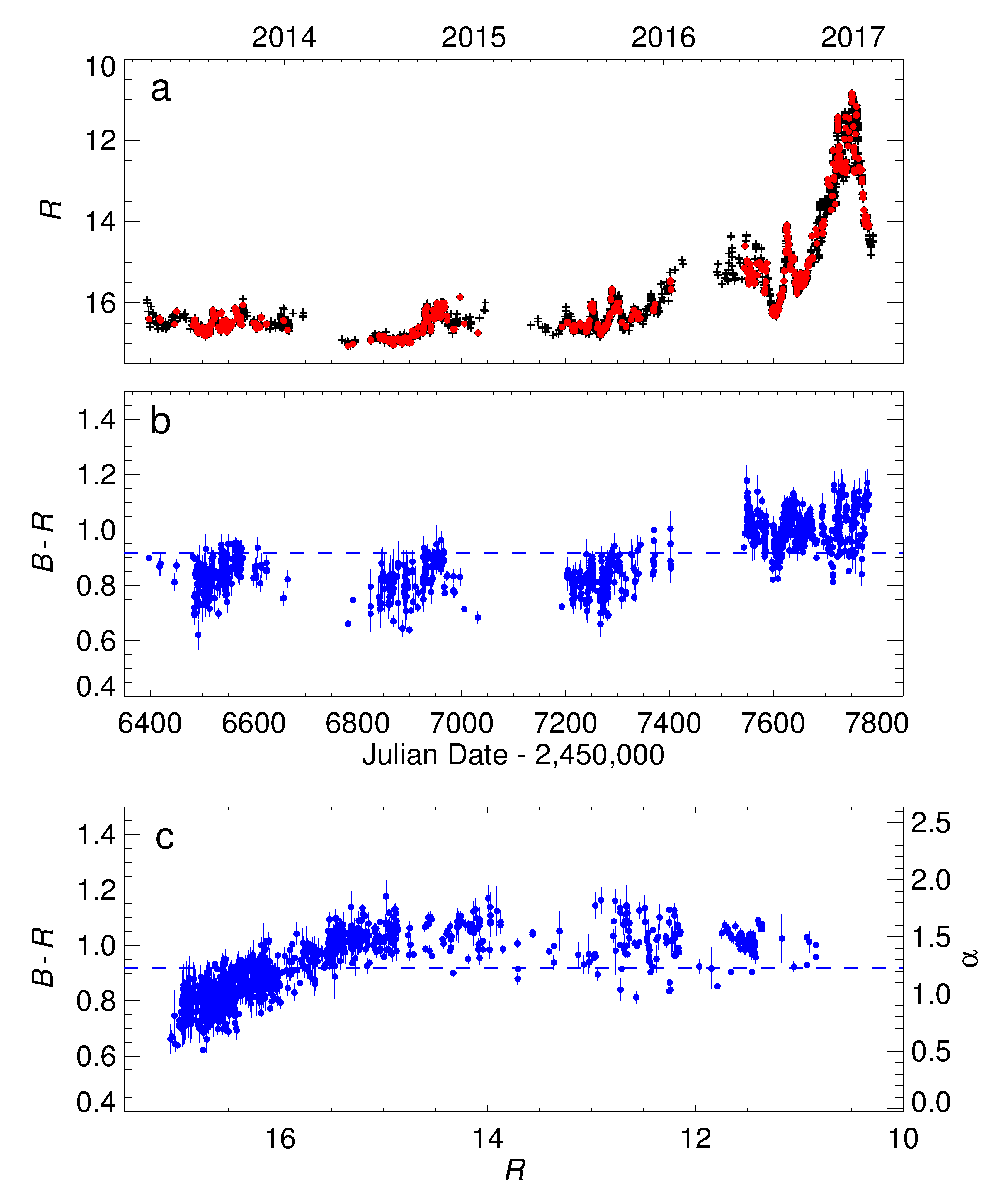}
\caption{{\bf Colour behaviour of CTA~102.} {\bf a}, The $R$-band light curve; red dots highlight the data points used to build colour indices. {\bf b-c}, The $B-R$ colour index as a function of time and of the $R$-band magnitude. Error bars on colours are obtained by summing in quadrature the 1 s.d. measure errors of the corresponding $B$ and $R$ data. The dashed line indicates the average $B-R$ value. The parameter $\alpha$ is the spectral index of the $F \propto \nu^{-\alpha}$ law.
The redder-when-brighter trend characterizing faint source states ($R > 15$) turns into a slight bluer-when-brighter trend as the source flux increases.}
\label{colori}
\end{figure}

\begin{figure}[h]
\center\includegraphics[width=10cm]{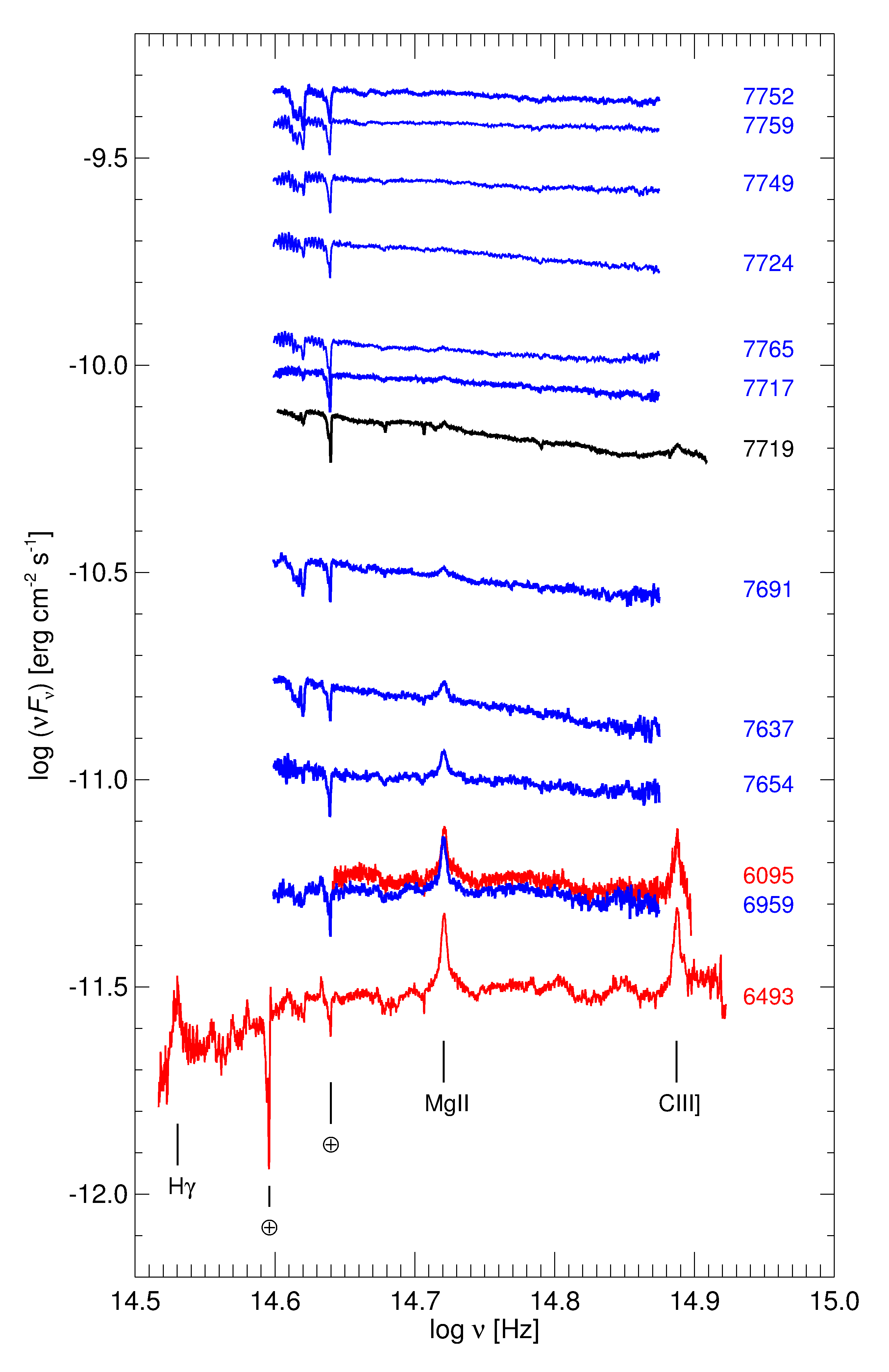}
\caption{{\bf A selection of optical spectra of CTA~102 in different brightness states.} Data are from the Steward (blue) and Roque de los Muchachos (TNG-black and NOT-red) observatories and have been corrected for Galactic extinction. Observing epochs are given on the right as $\rm Julian \, Date-2,450,000$. The main broad emission lines (more visible in faint states) are indicated. As the flux increases the source spectrum first softens (redder-when-brighter trend) and then gradually hardens (bluer-when-brighter).}
\label{spettri}
\end{figure}

\begin{figure}[h]
\center\includegraphics[width=\linewidth]{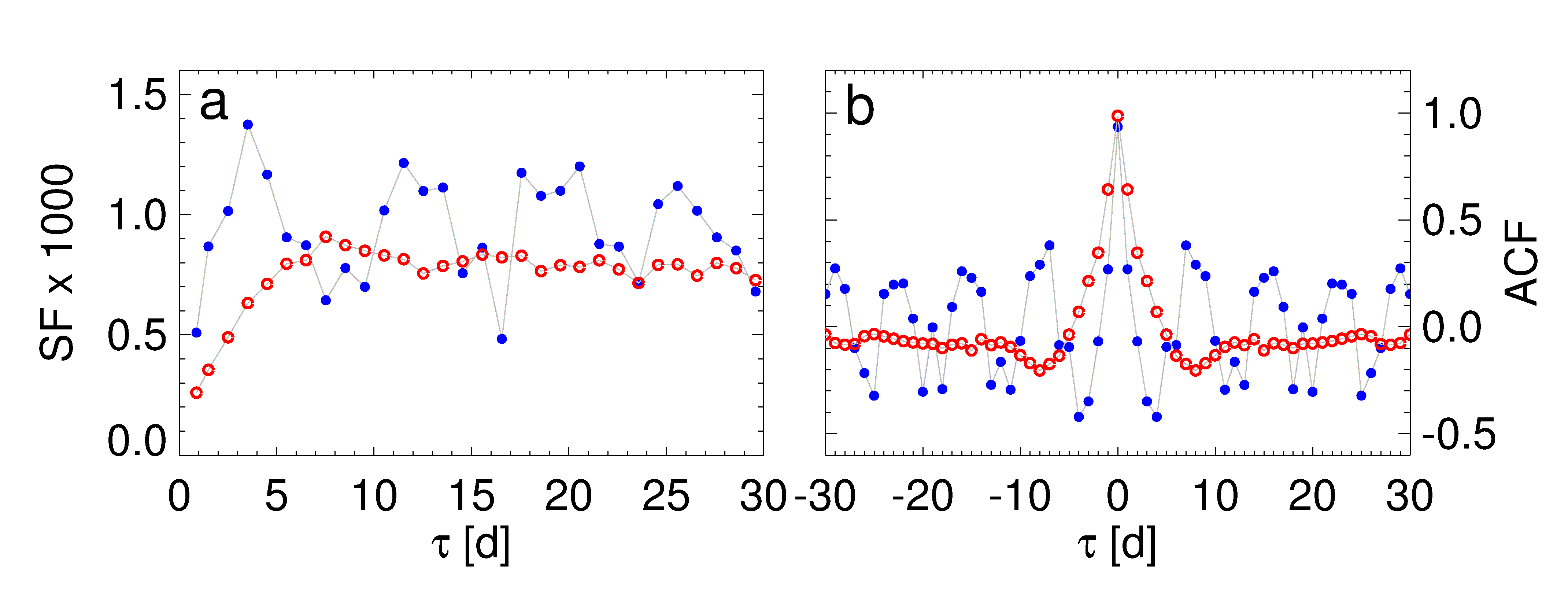}
\caption{{\bf Results of time series analysis on the optical fluxes.} {\bf a}, Structure function on the $R$-band flux densities corrected for the long-term trend due to variable relativistic beaming (see Fig.\ \ref{radop}). {\bf b}, Autocorrelation function on the same corrected fluxes. $\tau$ is the time separation between points, in bins of 1 day.
Filled blue and empty red symbols refer to bright (more beamed) and faint (less beamed) observed states, respectively, and show that variability time scales are halved when the Doppler factor doubles. }
\label{tisca}
\end{figure}

\begin{figure}[h]
\center\includegraphics[width=\linewidth]{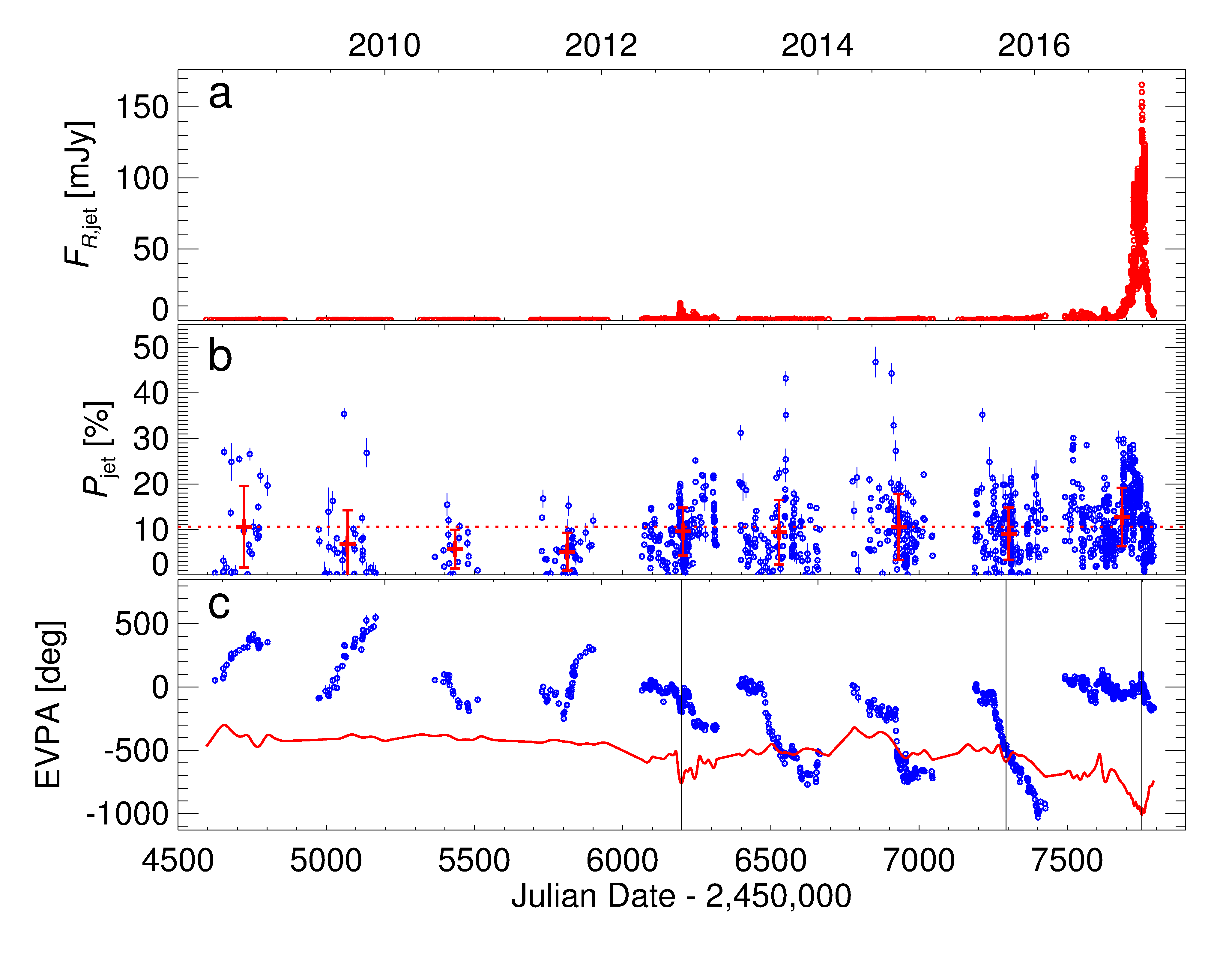}
\caption{{\bf Temporal behaviour of the CTA~102 polarization.} {\bf a}, The jet optical flux densities.
{\bf b}, The jet polarization percentage $P_{\rm jet}$. The horizontal dotted line indicates the average value over the whole period, while crosses show the mean values in each observing season. Error bars represent 1 s.d.
{\bf c}, The electric vector polarization angle EVPA. The red solid line displays the trend of the viewing angle in the $R$ band (see Fig.\ \ref{radop}) properly rescaled, and the vertical lines mark the most interesting events discussed in the text.
}
\label{pola}
\end{figure}

\begin{figure}[h]
\center\includegraphics[width=\linewidth]{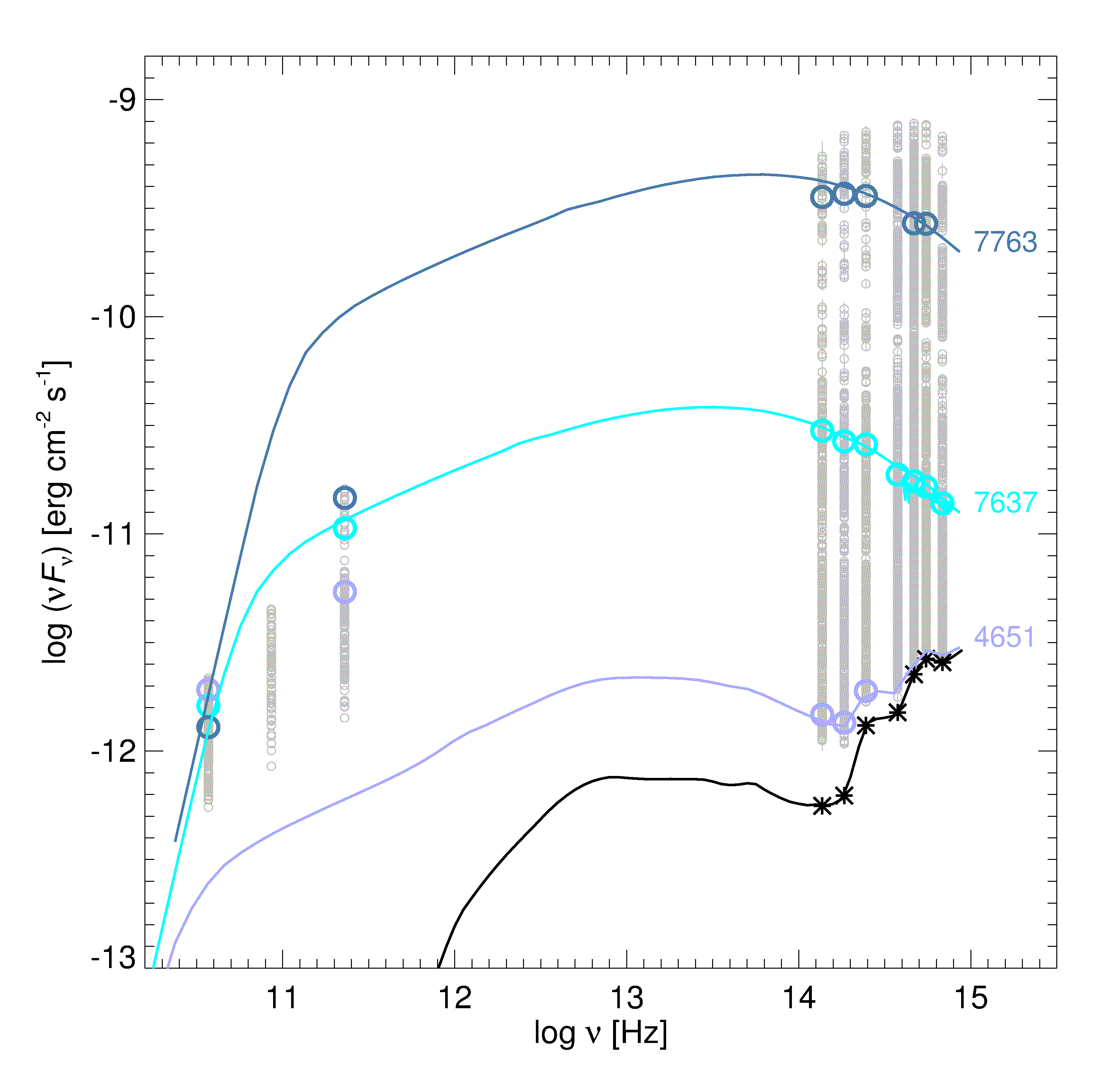}
\caption{{\bf One-zone model fits to spectral energy distributions of CTA~102.}
The standard one-zone model by [\ref{tra09}] has been used to fit three SEDs in an intermediate, high and low brightness state (see also Fig.\ \ref{sed}). Once the blob physical parameters are fixed to reproduce the intermediate state, the other two model fits are obtained by changing only the Doppler factor to match the optical data. As a result, the mm flux is largely over- or under-produced. In all model fits the thermal component (accretion disc+torus; black line and symbols) was added to the one-zone model synchrotron component.}
\label{one}
\end{figure}

\end{document}